\documentclass[useAMS,usenatbib,a4paper]{mn2e}
\usepackage{graphicx}
\usepackage{amsmath}
\usepackage{amssymb}

\title[Observable consequences of kinetic and thermal AGN
  feedback]{Observable consequences of kinetic and thermal AGN
  feedback in elliptical galaxies and galaxy clusters}

\author[E.C.D. Pope] {Edward
  C.D. Pope$^{1}$\thanks{E-mail:ecdpope@uvic.ca}\\ $^{1}$School of
  Physics \& Astronomy, University of Victoria, Victoria, BC, V8P
  1A1, Canada\\}

\begin{document}

\pagerange{\pageref{firstpage}--\pageref{lastpage} \pubyear{2009}}

\maketitle

\label{firstpage}

\begin{abstract}
We have constructed an analytical model of AGN feedback and studied
its implications for elliptical galaxies and galaxy clusters. The
results show that momentum injection above a critical value will eject
material from low mass elliptical galaxies, and leads to an X-ray
luminosity, $L_{\rm X}$, that is $\propto$ $\sigma^{8-10}$, depending
on the AGN fuelling mechanism, where $\sigma$ is the velocity
dispersion of the hot gas. This result agrees well with both
observations and semi-analytic models. In more massive ellipticals and
clusters, AGN outflows quickly become buoyancy-dominated. This
necessarily means that heating by a central cluster AGN redistributes
the intracluster medium (ICM) such that the mass of hot gas, within
the cooling radius, should be $ \propto L_{\rm X}(<r_{\rm
  cool})/[g(r_{\rm cool})\sigma]$, where $g(r_{\rm cool})$ is the
gravitational acceleration at the cooling radius. This prediction is
confirmed using observations of seven clusters. The same mechanism
also defines a critical ICM cooling time of $\sim 0.5$ Gyr, which is
in reasonable agreement with recent observations showing that star
formation and AGN activity are triggered below a universal cooling
time threshold.
\end{abstract}


\section{Introduction}

It is well known that elliptical galaxies are commonly the hosts of
powerful AGN \citep[][]{mclure}. These sources give rise to lobes of
radio emission embedded in the X-ray emitting gaseous haloes
surrounding the galaxies and permeating clusters of galaxies
\citep[e.g.][]{burns,best05,birzan,dunn05,rafferty06,best07,nulsen2007,diehl08}.

Recent observational studies of AGN in elliptical galaxies, and galaxy
clusters, also suggest that AGN activity is related to the thermal
state of its environment. Systems with short radiative cooling times,
or a low central entropy, have been shown to be more likely to host
active star formation, optical line-emission and jet-producing AGN
\citep[e.g.][]{burns,crawf,raff08,cav08,mittal}. This suggests that
AGN activity is part of a feedback loop that is likely to have
important consequences for its environment.

The theoretical evidence for the importance of AGN feedback has also
continued to mount. \cite{tabor93} and \cite{bintab} employed AGN
feedback to prevent catastrophic radiative cooling of the X-ray
emitting gas that surrounds elliptical galaxies. If unopposed,
radiative cooling would lead to the deposition of large quantities of
cold gas in the central galaxy, and thus cause the galaxy to grow
rapidly. Since this appears not to be the case, it was assumed that
there must be a heat source that balances the cooling. In a different
approach to a similar problem, semi-analytic models of galaxy
formation and evolution \citep[e.g.][]{benson,croton05,bower06} also
demonstrated that AGN feedback was necessary to reproduce the observed
galaxy luminosity function at large masses, and balance the large
radiative cooling rates in clusters of galaxies. In the absence of AGN
feedback, the models produced too many massive galaxies. 

Semi-analytic models \citep[e.g.][]{benson}, hydrodynamic simulations
\citep[e.g.][]{bregdav88,gaetz,pope05} and observations
\citep[e.g.][]{voigt04,pope06} have all shown that thermal conduction
cannot provide the required energy transfer rates to balance cooling
in the majority of cases. However, numerous hydrodynamical simulations
have shown that a combination of AGN heating and thermal conduction
can maintain the system in a steady-state
\citep[e.g.][]{ruszbegel02,brueggen03,hoeft,roy04}, see also
\cite{guo} for a stability analysis.

Recent theoretical efforts have focussed on trying to understand the
X-ray luminosity/temperature ($L_{\rm X}-T$) relation in elliptical
galaxies and galaxy clusters. The results of semi-analytic models
\citep[e.g.][]{bower08, dave} and hydrodynamic models
\citep[e.g.][]{puchwein} have shown that AGN feedback does indeed play
a significant role in governing the scaling of this relation.

The aim of this article is to describe the impact on elliptical
galaxies of momentum and thermal energy injection by a central
AGN. There are critical thresholds above which each energy component
will eject material from the galaxy, and so each mechanism leads to
clear observable signatures which can be used to distinguish between
them. This study is restricted to the study of galaxies where the
X-ray emission is dominated by the gaseous atmosphere, rather than
stellar sources \citep[e.g.][]{os01}.

To this end, we have employed an analytical model of AGN feedback in
order to better understand, and compliment, the results that have
emerged from semi-analytic and hydrodynamic simulations, as well as
observations.

The article is arranged as follows: in section 2 we introduce the
basic model describing the effect of energy and momentum injection
into a hot gaseous atmosphere. Section 3 contains the results for the
limiting cases of momentum dominated and thermally dominated outflows
and discusses the implications for the $L_{\rm X}-\sigma$ relation in
elliptical galaxies. Note, we have chosen to study the $L_{\rm
  X}-\sigma$ relation rather than $L_{\rm X}-T$, because the
gravitational potential can be written in terms of $\sigma$ which we
assume to be unaltered by feedback. In section 4, the theory is
applied directly to observations of galaxy clusters. The main findings
are summarised in section 5.

\section{The Model}

The aim of this section is to develop a simple, but useful, analytical
formulation that can describe the general impact of AGN feedback on
the $L_{\rm X}-\sigma$ relation of elliptical galaxies. For this
purpose the elliptical galaxy is approximated as a singular isothermal
sphere \citep[e.g.][]{ka03,ka05,murray05}. This is because the setup
has proved to be a successful starting point in describing phenomena
such as the black hole mass/stellar velocity dispersion relation. The
results are used to illustrate general behavioural trends, and are not
intended to be a complete description of all the physical processes
involved. However, in Section 4, which focuses on galaxy clusters,
observational data are used to calculate more accurate gravitational
potentials and gas masses. This allows specific predictions of the
model to be tested rigorously on individual systems.

For the singular isothermal sphere, the mass of gas within a radius R,
is
\begin{equation}
M_{\rm T}(R) = 4\pi \int_{0}^{R} \rho r^{2}{\rm d}r = \frac{2 f_{\rm
    g}\sigma^{2}R}{G},
\end{equation}
where $f_{\rm g} \approx 0.17$ \citep[][]{spergel07} is the gas
fraction, $\sigma$ is the velocity dispersion, $R$ is the radius, and
$G$ is Newton's gravitational constant. In this model, the AGN resides
at the centre of the potential, and produces outflows that propagate
radially outwards. The mass swept up by outflows subtending a solid
opening angle, $\Omega$, is therefore
\begin{equation}\label{eq:sw}
M(R) = \bigg(\frac{\Omega}{4\pi}\bigg)\frac{2 f_{\rm
    g}\sigma^{2}R}{G},
\end{equation}
so that $M(R) = \Omega M_{\rm T}(R)/(4\pi)$. As long as the outflow is
active, new material from the gaseous atmosphere will be prevented
from entering this solid angle through the sides. Therefore,
equation(\ref{eq:sw}) gives a reasonable estimate of the mass
collected by the outflow. The darkmatter gravitational potential is
assumed to be unaffected by the movement of the gas.

Considering momentum injection at a rate of $\dot{P}$, Newton's 2nd
law gives
\begin{equation}\label{eq:newt}
\frac{{\rm d}[M(R)\dot{R}]}{{\rm d}t} = \dot{P} + g(R)M(R),
\end{equation}
where $g(R)$ is the gravitational acceleration. If $\dot{P} \ge
-g(R)M(R)$, momentum injection will overcome the gravity of the galaxy
and the AGN will induce a bulk outflow of hot material collected from
the galaxy's atmosphere.

However, the outflow must do work against both gravity and against the
pressure of the gaseous atmosphere it is lifting. This means that a
term, $(\Omega/4\pi)4\pi R^{2}p_{\rm amb}$, must be subtracted from
the right hand side of equation (\ref{eq:newt}), where $p_{\rm amb}$
is the ambient pressure. Using the ideal gas equation, the pressure
term can be written
\begin{equation}
\bigg(\frac{\Omega}{4\pi}\bigg)4\pi R^{2}p_{\rm amb}(R) =
\bigg(\frac{\Omega}{4\pi}\bigg)4\pi R^{2}\frac{\rho(R)k_{\rm b}T}{\mu
  m_{\rm p}} = \bigg(\frac{\Omega}{4\pi}\bigg)\frac{2f_{\rm
    g}\sigma^{4}}{G},
\end{equation}
where $k_{\rm b}T/\mu m_{\rm p} \equiv \sigma^{2}$. 

The AGN also injects thermal energy, at a rate $\dot{E}_{\rm
  thermal}$, that will exert an outward pressure on the ambient
gas. Assuming that the inflation is reversible
\begin{equation}
\dot{E}_{\rm thermal} =
\frac{\gamma}{\gamma-1}\bigg(\frac{\Omega}{4\pi}\bigg)4\pi
R^{2}\dot{R}p_{\rm amb}(R),
\end{equation}
where $\gamma$ is the adiabatic index of the injected gas. Therefore,
we must add a term
\begin{equation}
\bigg(\frac{\Omega}{4\pi}\bigg)4\pi R^{2}p_{\rm amb}(R) =
\frac{\gamma-1}{\gamma}\frac{\dot{E}_{\rm thermal}}{\dot{R}},
\end{equation}
to the right hand side of equation (\ref{eq:newt}).\citep[This term
  should be compared with the $L_{\rm Eddington}/c$ term used in][to
  model the outflow momentum.]{ka03,ka05} The completed equation of
motion can be written
\begin{equation} \label{eq:total}
\frac{{\rm d}[M(R)\dot{R}]}{{\rm d}t} = \dot{P} + g(R)M(R)
-\bigg(\frac{\Omega}{4\pi}\bigg)\frac{2f_{\rm g}\sigma^{4}}{G} +
\frac{\gamma-1}{\gamma}\frac{\dot{E}_{\rm thermal}}{\dot{R}}.
\end{equation}

\section{Results}

In this section we investigate the behaviour of equation
(\ref{eq:total}) in two limits: momentum-dominated AGN heating, and
thermally-dominated AGN heating.

\subsection{Kinetic feedback}

In the limit that the outflow is completely momentum driven, the
critical value of the momentum injection rate for the singular
isothermal sphere is
\begin{equation}\label{eq:kin}
\dot{P}_{\rm crit} = -g(R)M(R)
+\bigg(\frac{\Omega}{4\pi}\bigg)\frac{2f_{\rm g}\sigma^{4}}{G}=
\bigg(\frac{\Omega}{4\pi}\bigg)\frac{6f_{\rm g}\sigma^{4}}{G}.
\end{equation}
In general, it is more intuitive to consider the energy injection rate
rather than the momentum flux. Therefore, equation (\ref{eq:kin})
shall be re-written accordingly. The momentum injection rate is
related to the kinetic energy injection rate by
\begin{equation} \label{eq:mom}
\dot{E}_{\rm kinetic} = \frac{\dot{P}^{2}}{2\dot{m}_{\rm out}},
\end{equation}
where $\dot{m}_{\rm out}$ is the mass flow rate of the jet.

Using equation (\ref{eq:mom}) with equation (\ref{eq:kin}) and
rearranging for the critical luminosity gives
\begin{equation}
\dot{E}_{\rm kinetic,crit} = \frac{9\Omega^2 \sigma^{8}f_{\rm
    g}^{2}}{8\pi^{2} G^{2}\dot{m}_{\rm out}},
\end{equation}
which, in scaled units, is
\begin{eqnarray}\label{eq:crit}
\dot{E}_{\rm kinetic, crit} \sim 2 \times 10^{42}
\bigg(\frac{\Omega/4\pi}{0.02}\bigg)^{2}\bigg(\frac{\dot{m}_{\rm
    out}}{10\,{\rm M_\odot}{\rm
    yr^{-1}}}\bigg)^{-1}\nonumber\\ \bigg(\frac{\sigma}{200\,{\rm
    km\,s^{-1}}}\bigg)^{8}\,{\rm erg\,s^{-1}}.
\end{eqnarray}
The reason $\Omega/4\pi = 0.02$ has been chosen will become apparent
in section 4, where the value is derived from observations. At
present, it is sufficient to note that $\Omega$ is an effective
opening angle, rather than the actual opening angle of the outflow. It
is the solid angle within which the outflow interacts with ambient
material, and is therefore larger than the typical value dervied from
radio observations \citep[e.g. see][and references therein]{bintab}.

The luminosities hinted at in equation (\ref{eq:crit}) mean that it is
plausible to expect kinetic AGN feedback to eject hot material from
the atmosphere of the galaxy. This will have two effects: it will
reduce the X-ray luminosity of the galaxy, $L_{\rm X}$, and reduce the
gas fraction. 

In clusters (with large $\sigma$), the critical jet power will exceed
the cluster X-ray luminosity, and would require implausibly large jet
powers to eject mass. This means that the AGN jets are unlikely to
significantly affect the $L_{\rm X} \propto \sigma^{4.4}$ relation
\citep[][]{mahdavi} in clusters. Of course, this does not mean that
they do not play a significant role in heating the inner regions of
such objects.

To make a realistic estimate of the $L_{\rm X}-\sigma$ relation
expected as a consequence of momentum-dominated feedback, we follow
the method of \cite{bower08}. Their prescription for the effect of the
AGN activity was to remove mass from the galaxy at a rate governed by
the difference between the heating and cooling rates: $\dot{m} =
2(\dot{E}-L_{\rm X})/\sigma^{2}$. Our model is expected to behave in a
similar way, but with the reference point being the critical jet power
defined in equation (\ref{eq:crit})
\begin{equation} \label{eq:semi}
\dot{m} = \frac{2(\dot{E}_{\rm kinetic}-\dot{E}_{\rm
    crit})}{\sigma^{2}}.
\end{equation}
Equation (\ref{eq:semi}) tells us that the mass-loss from the galaxy
will effectively stop when the average feedback heating rate becomes
comparable with the critical jet power for that gravitational
potential.

The complete derivation requries a functional form for $\dot{E}_{\rm
  kinetic}$. \cite{bower08} used an energy injection rate that is
proportional to the mass cooling rate, $\dot{M}_{\rm cool}$, of the
X-ray emitting atmosphere around the galaxy. We will follow their
example as an illustrative case. Therefore
\begin{equation}
\dot{E}_{\rm kinetic} = \eta \delta \dot{M}_{\rm cool}c^{2},
\end{equation}
where the factor $\delta$ accounts for the fraction of cooling
material that reaches the black hole. Here, $\delta$ is also used to
account for the reduced cooling rate compared to the classical
value. Given that cooling in such systems appears to be reduced by a
factor of 10, and assuming that only one tenth of the inflowing
material reaches the black hole, $\delta \sim 0.01$. $\eta$ is the
accretion efficiency, and $c^{2}$ is the speed of light. The classical
mass cooling rate can be written in terms of the X-ray luminosity of
the hot gas, as
\begin{equation}
\dot{M}_{\rm cool} = \frac{2}{5}\frac{L_{\rm X}}{\sigma^{2}},
\end{equation}
so that the kinetic energy injection rate is
\begin{equation}\label{eq:edot}
\dot{E}_{\rm kinetic} = 0.4 \eta \delta L_{\rm X}
\frac{c^{2}}{\sigma^{2}}.
\end{equation}
Substituting equation (\ref{eq:edot}) into equation (\ref{eq:semi})
shows that the mass-loss rate from the galaxy will tend to zero as
\begin{equation}
L_{\rm X} \sim 2.5\frac{\sigma^{2}\dot{E}_{\rm kinetic, crit}}{\eta
  \delta c^{2}} \propto \sigma ^{10}.
\end{equation}
In scaled units, this means we expect the X-ray luminosity of an
elliptical galaxy to be approximately
\begin{eqnarray}\label{eq:lx}
L_{\rm X} \sim 2 \times
10^{39}\bigg(\frac{\Omega/4\pi}{0.02}\bigg)^{2}\bigg(\frac{\delta}{0.01}\bigg)^{-1}
\bigg(\frac{\dot{m}_{\rm out}}{10\,{\rm M_\odot}{\rm
    yr^{-1}}}\bigg)^{-1}\nonumber\\ \bigg(\frac{\sigma}{200\,{\rm
    km\,s^{-1}}}\bigg)^{10}\,{\rm erg\,s^{-1}},
\end{eqnarray}
where an accretion efficiency of $\eta = 0.1$ has been assumed. These
values agree reasonably well with the observed X-ray luminosities of
elliptical galaxies \citep[e.g.][]{fukazawa}. The result agrees well
with the observations presented by \cite{mahdavi} who also found
$L_{\rm X} \propto \sigma^{10}$, and is consistent with the relation
found by \cite{osullivan03}. The normalisation of the relation found
by \cite{mahdavi} seems to be slightly higher than equation
(\ref{eq:lx}), possibly suggesting that $\delta \sim 10^{-3}$ would be
a better description of the physics governing the accretion rate. This
would guarantee a low Eddington fraction accretion rate which would
produce outflows \citep[e.g.][]{macc} and not quasars.

The scaling relation in equation (\ref{eq:lx}) is steeper than the
results of numerical simulations performed by \cite{dave} who found
$L_{\rm X} \propto \sigma^{8.2}$. However, our model would also
produce $L_{\rm X} \propto \sigma^{8}$ if the AGN fuelling rate was
proportional to $L_{\rm X}$ rather than $L_{\rm X}/\sigma^{2}$. In
both observations and simulations, there is significant scatter in
these relationship. Presumeably this could be caused by the different
histories of individual galaxies (mergers and ages) and also the
stochastic nature of AGN feedback.

\subsection{Thermal energy injection}

It is also important to consider the role of thermal energy
injection. According to equation (\ref{eq:total}), the effect of
thermal injection should be quite different to pure momentum
injection. This leads to several predictions which can be tested
against existing observations.

A variant of this mechanism is generally considered to lead to the
black hole mass - bulge stellar velocity dispersion relation
\citep[e.g.][]{puchwein}. However, in their approach the radiative
power output of the AGN accretion at a high Eddington ratio couples to
the ambient gas. We will not be considering effects of radiation
output from the AGN since it is assumed here that the AGN simply
injects thermal energy that does work against its surroundings, rather
than assuming anything more complex.

Using equation (\ref{eq:total}), the critical energy injection rate
for the thermal gas to eject material from the gravitational potential
is of the order
\begin{equation}
\dot{E}_{\rm thermal, crit} \sim
\bigg(\frac{\Omega}{4\pi}\bigg)\frac{6f_{\rm g}\sigma^{5}}{G},
\end{equation}
where we have set $\dot{R} \sim \sigma$. In scaled units this gives
\begin{equation}
\dot{E}_{\rm thermal, crit} \sim 10^{42}
\bigg(\frac{\Omega/4\pi}{0.02}\bigg)\bigg(\frac{\sigma}{200\,{\rm
    km\,s^{-1}}}\bigg)^{5}\,{\rm erg\,s^{-1}}.
\end{equation}
In reality, the AGN injects both momentum and thermal energy (as well
as the radiation that we are not considering), rather than one or the
other. The behaviour of the system will therefore depend on the
relative proportions.

For simplicity, the thermal energy injection rate can be written as a
fraction, $\chi$, of the total energy injection rate; $\dot{E}_{\rm
  thermal} = \chi\dot{E}_{\rm total}$. Therefore, the critical
threshold for thermal energy injection, in terms of the total energy
injection rate, is $\dot{E}_{\rm total, crit} = \dot{E}_{\rm thermal,
  crit}/\chi$. It is probably safe to assume that initially, at least,
outflows are momentum dominated, since they are highly
anisotropic. This suggests that $\chi \ll 1$. In the next section we
will show that, above a critical jet length, AGN outflows will become
buoyancy dominated so that $\chi \rightarrow 1$ for large
radii. However, for the remainder of this section it is assumed that
the relative fractions of kinetic and thermal energy in the outflow
remain constant, and that the outflows are momentum dominated.

Combining the critical luminosities for both thermal and kinetic
limits we find that the critical energy injection rate can be
expressed as a broken power-law that scales as $\sigma^{8}$ at lower
masses, and $\sigma^{5}$ at higher masses. The location of the break
depends on the value of $\chi$; $(\sigma_{\rm break}/200\,{\rm
  km\,s^{-1}}) \sim (2\chi)^{-1/3}$. So, as an example, if $\chi=
0.1$, the break will occur at $\sigma \sim 340\,{\rm
  km\,s^{-1}}$. Above this value of $\sigma$, the criterion for mass
ejection from the galaxy follows the thermal limit, while below this
the mass ejection criterion is governed by momentum injection. Another
way of expressing this is that thermal energy injection has much more
important consequences for the galaxy properties at high masses, while
momentum injection is more important at lower masses. This result is
shown in figure 1.

\begin{figure}
\centering
\includegraphics[width=10cm]{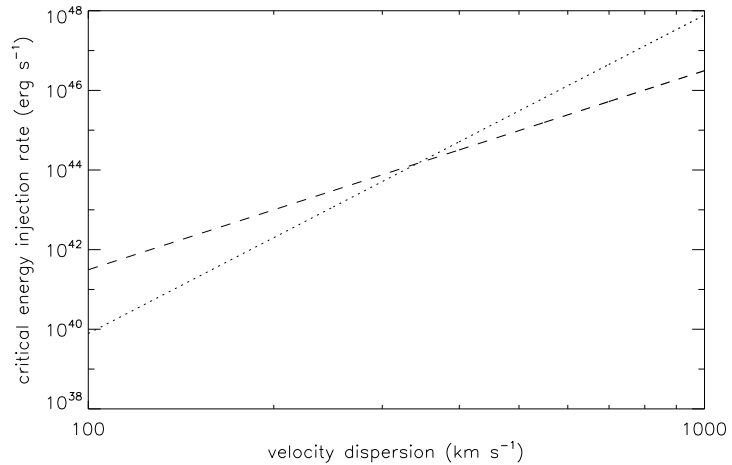}
\caption{The critical energy injection rate for thermal (dashed line)
  and momentum (dotted line) injection. In this example it was assumed
  that the thermal fraction, $\chi=0.1$ so that the critical values
  cross at roughly $340\,{\rm km\,s^{-1}}$. At low $\sigma$, the lower
  critical energy injection limit is provided by the momentum
  criterion, while at high $\sigma$ the lower critical injection limit
  is provided by thermal energy criterion.}
\label{fig:flow2}
\end{figure}

The critical energy injection rate at $\sigma \sim 340\,{\rm
  km\,s^{-1}}$ is $\sim 10^{44}\,{\rm erg\,s^{-1}}$ which is well
below the Eddington limit for a $10^{8}\,M_{\odot}$ black hole. This
strongly suggests that thermal feedback from black holes could have an
effect on the $L_{\rm X}-\sigma$ relation in massive elliptical
galaxies and groups. However, for a cluster gravitational potential,
$\sigma > 500\,{\rm km\,s^{-1}}$. The critical energy injection rate
at $\sigma = 500\,{\rm km\,s^{-1}}$, is $\sim 10^{47}\,{\rm
  erg\,s^{-1}}$, which is comparable to the Eddington limit for a
$10^{9}\,M_{\odot}$ black hole. Furthermore, depending on the value of
$\delta$ the actual power available from accretion might be much lower
than this. Consequently, there should be a second break in the $L_{\rm
  X}-\sigma$ relation at $\sigma \sim 500\,{\rm km\,s^{-1}}$. Assuming
that $\sigma^{2} = k_{\rm b}T/\mu m_{\rm p}$, $\sigma = 500\,{\rm
  km\,s^{-1}}$ corresponds to a temperature of $\sim 1-2$ keV, which
is approximately where a break is observed \citep[e.g.][]{xue}. This,
therefore, marks the transition above which black holes are unlikely
to affect the global $L_{\rm X}-\sigma$ relation. Nevertheless, black
holes may still be effective in heating the central regions of
clusters.

If the accretion rate onto the black hole is $\propto L_{\rm
  X}/\sigma^{2}$, as suggested in the previous example, thermal AGN
feedback should lead to $L_{\rm X} \propto \sigma^{7}$ in higher mass
ellipticals and groups. Alternatively, if the black hole accretion
rate is $\propto L_{\rm X}$, the model would lead to $L_{\rm X}
\propto \sigma^{5}$. 

So, in summary, the model predicts $L_{\rm X} \propto \sigma^{8-10}$
for lower masses elliptical galaxies, $L_{\rm X} \propto \sigma^{5-7}$
for higher mass ellipticals and groups, depending on the AGN fuelling
mechanism. However, given that the relation for clusters can be
written $L_{\rm X} \propto \sigma^{4.4}$ \citep[][]{mahdavi} and there
is a noticeable break in the scaling at $\sim 1-2$ keV
\citep[e.g.][]{xue} (or $\sigma \sim 500\,{\rm km\,s^{-1}}$ ), it
seems that an AGN fuelling rate that is $\propto L_{\rm X}/\sigma^{2}$
can best explain the observed $L_{\rm X}-\sigma$ relation across the
mass range considered here.

\section{Application to galaxy clusters}

Having considered the implications of momentum and thermal energy
injection on elliptical galaxies, the model is applied to clusters of
galaxies. Because of the stronger gravity, these systems behave rather
differently to elliptical galaxies. This leads to several observable
consequences which are discussed below. In addition, the benefit of
using the singular isothermal sphere for elliptical galaxies was to
construct simple scaling laws across the population of galaxies that
result from different properties of AGN feedback. However, the
previous section demonstrated that AGN feedback is unable to alter the
global $L_{\rm X}-\sigma$ relation in clusters, so using the same
general approach would not yield interesting results for
clusters. Instead, it seems more appropriate to focus on the central
regions of clusters, where the AGN is likely to have an impact and so
attempt to discern any observable signatures of AGN feedback. To do
this accurately requires observations of the temperature and density
profiles of the central ICM in a sample of clusters. This is done in
Section 4.2. Firstly, though, there is brief discussion of the general
effect of the cluster potential on AGN outflows.

\subsection{Buoyancy, work and bubbles in the intracluster medium}

Momentum and thermal energy fractions govern various properties of the
outflow, including its morphology. These fractions are not constant,
but vary as the outflow propagates through its surroundings. For
example, thermal energy injection produces buoyant material, and the
buoyancy force is also a function of the local gravitational
acceleration. This suggests that identical jets will appear
morphologically different in different environments. In particular,
buoyancy is more important in more massive ellipticals, and in
clusters, due to the stronger gravity in such enviroments. It,
therefore, stands to reason that we might expect outflows to appear
more bubble-like (buoyant), and less elongated, in clusters. This
possibility is investigated below.

For a given thermal energy injection rate, it is possible to calculate
a buoyancy flux. An outflow of velocity, $w$, and width, $D$, has a
buoyancy flux defined as \citep[e.g.][and references
  therein]{carlotti}
\begin{equation}
B \equiv \frac{\Delta \rho}{\rho}g w \frac{\pi D^{2}}{4}\,{\rm
  cm^{4}\,s^{-3}},
\end{equation}
where $g$ is the gravitational acceleration, $\rho$ is the ambient
density, and $\Delta \rho$ is the density difference between the
outflow and the ambient material. The buoyancy flux can also be
written in terms the heating rate that caused the density change
\citep[e.g.][]{carlotti}
\begin{equation}\label{eq:b}
B = \frac{\gamma-1}{\gamma}\frac{g\dot{E}_{\rm thermal}}{p_{\rm
    amb}}\,{\rm cm^{4}\,s^{-3}},
\end{equation}
where $p_{\rm amb}$ is the pressure of the ambient gas. Similarly, the
momentum flux is defined as \citep[e.g.][]{carlotti}
\begin{equation}
U \equiv \bigg(\frac{\rho_{\rm outflow}}{\rho}\bigg)w^{2} \frac{\pi
  D^{2}}{4} = \frac{\dot{P}}{\rho}\,{\rm cm^{4}\,s^{-2}},
\end{equation}
where $\rho_{\rm outflow}/\rho$ is the ratio of the density in the jet
to the ambient density.

It is useful to construct a characteristic length-scale from the
momentum and buoyancy fluxes. This length-scale marks the transition
from a momentum-dominated flow to a buoyancy-dominated flow and is
defined as
\begin{equation}
l \equiv \frac{U^{3/4}}{B^{1/2}}.
\end{equation}
Using the fact that $\dot{P} = (2\dot{m}_{\rm out}\dot{E}_{\rm
  kinetic})^{1/2}$, the length-scale can be written
\begin{eqnarray}\label{eq:l}
l \sim 24\,{\rm kpc} \bigg(\frac{\dot{m}_{\rm out}}{10\,{\rm M_{\odot}
    yr^{-1}}}\bigg)^{3/8}\bigg(\frac{\dot{E}_{\rm
    kinetic}}{10^{42}\,{\rm
    erg\,s^{-1}}}\bigg)^{3/8}\nonumber\\ \bigg(\frac{\rho}{10^{-25}\,{\rm
    g\,cm^{-3}}}\bigg)^{-3/4}\bigg(\frac{p_{\rm amb}}{10^{-10}\,{\rm
    erg\,cm^{-3}}}\bigg)^{1/2}\bigg(\frac{\dot{E}_{\rm
    thermal}}{10^{42}\,{\rm
    erg\,s^{-1}}}\bigg)^{-1/2}\nonumber\\\bigg(\frac{g}{10^{-8}\,{\rm
    cm s^{-2}}}\bigg)^{-1/2}.
\end{eqnarray}
We have assumed a uniform density for the environment, and constant
gravity, but nevertheless, the critical length above which the jet
dynamics become dominated by buoyancy is likely to be $\sim $ 20
kpc. 

The parameters that control this length-scale show that an outflow
will become buoyancy-dominated closer to its origin in a denser
environment with stronger gravity. This confirms the earlier statement
that, for a given energy injection rate, bubble-like outflows will be
more prevalent in the centres of clusters of galaxies than in low mass
elliptical galaxies. It also means that more of the injected energy is
trapped within the central regions, providing more efficient heating
in a cluster environment.

In clusters, it is not possible to relate many of the properties, such
as density and pressure, to $\sigma$. As a result, it is not
particularly useful to represent equation (\ref{eq:l}) in terms of the
velocity dispersion. In addition, there is no single value of $\sigma$
when the jets become buoyant, because this will depend on the kinetic
and thermal energy injection rates and the other cluster properties.

Another possible explanation for the presence of AGN-blown bubble-like
structures in clusters is that jets never exceed the critical energy
injection to eject mass from a cluster. Therefore, the jet will always
be constrained by the ambient medium. In contrast, it is possible for
the AGN to exceed the critical luminosity in an elliptical galaxy, in
which case the jet is no longer constrained. It is possible that this
effect plays some role in the apparent FRI/FRII dichotomy. For
example, \cite{kb07} suggested that FRIs were those objects which are
disrupted within the core region the surrounding atmosphere, while
FRIIs were jets which were sufficiently powerful to penetrate into the
region outside the core where the ambient pressure drops off
rapidly. Note, however, other effects such as black hole spin are also
a possible explanation for the dichotomy \citep[e.g.][]{meier99}.

It is also necessary to show that the term $\dot{E}_{\rm
  thermal}/\dot{R}$ term in equation (\ref{eq:total}) is equivalent to
the work done by bouyancy against the surrounding gas. The rate of
change of the buoyancy force, according to Archimedes' principle, can
be written
\begin{equation}\label{eq:buo}
\dot{F} = \bigg(\frac{\Omega}{4\pi}\bigg)4\pi \rho g R^{2}\dot{R},
\end{equation}
where the density of the outflow is assumed to be much less than the
material it has displaced. Using equation (\ref{eq:b}) the rate of
change buoyancy force in terms of the heating rate, $\dot{E}_{\rm
  thermal}$, is
\begin{equation}\label{eq:buo2}
\dot{F} = \frac{\gamma-1}{\gamma}\frac{\mu m_{\rm p}g \dot{E}_{\rm
    thermal}}{k_{\rm b}T}.
\end{equation}
Equations (\ref{eq:buo}) and (\ref{eq:buo2}) can then be rearranged to
show that the two forms are equivalent
\begin{equation}
\frac{\gamma-1}{\gamma}\frac{\dot{E}_{\rm thermal}}{\dot{R}} =
\bigg(\frac{\Omega}{4\pi}\bigg)4\pi p_{\rm amb} R^{2}.
\end{equation}
This means that buoyancy has the same effect as thermal energy
injection doing work against its surroundings. Furthermore, since the
outflows become buoyancy-dominated on length-scales greater than $l$,
this means that the initial momentum flux is converted into thermal
energy. The total power must be conserved, so for $R>l$, the outflow
can be adequately described by $\dot{E}_{\rm thermal} \sim
\dot{E}_{\rm total}$, where $\dot{E}_{\rm total}$ is the (constant)
total energy injection rate of the outflow. As a result, the factor
$\chi$, introduced earlier, is a strong function of radius and becomes
close to unity outside $R=l$.

Equation (\ref{eq:l}) suggests that $l$ is likely to be significantly
less than the cooling radius in a typical cluster ($\sim$ 100
kpc). Therefore, we argue that on spatial scales comparable with the
cooling radius, it is sufficient to model heating by an AGN at the
cluster centre, purely in terms of the injection of thermal energy
within a given solid angle. To understand the dynamical structures on
smaller scales we must consider the jet momentum in detail, but on
intermediate spatial scales it seems that the energy injection can be
treated as purely thermal. We will next consider the physical
implications of thermal heating in galaxy clusters.

\subsection{Observational consequences of buoyancy in cluster cores}

In clusters, the AGN cannot eject gas from the cluster due to the
sheer weight of material residing in the gravitational
potential. However, there is still a region near the cluster centre
where the AGN can affect the distribution of the X-ray emitting
material. Since we are considering clusters, buoyancy is important and
the energy injection can be thought of as purely thermal. Therefore,
the appropriate critical energy injection rate is the thermal
injection limit; $\dot{E}_{\rm crit} \sim -(15/4)\sigma M(R)g(R)$. As
usual, $M(R)$ is the mass of the material swept up within $R$ and
within the solid angle of the outflow; $M(R) = \Omega \int_{0}^{R}
r^{2}\rho(r){\rm d}r = \Omega M_{\rm T}(R)/(4\pi)$. $\sigma$ is the
velocity dispersion of the gravitational potential, and $g(R)$ is the
gravitational acceleration at $R$. In general, clusters are close to
hydrostatic equilibrium so that the gravitational acceleration can be
calculated from the pressure gradient and the gas density; $g(R)
\approx \nabla p (R) / \rho(R)$. The multiplicative factor $15/4$
comes from $(3/2) \times \gamma/(\gamma-1)$, assuming that $\gamma =
5/3$. Finally, the factor $3/2$ is from the contribution of the
ambient pressure to the total work required for the thermal outflow to
eject material, see equation (\ref{eq:total}).

For energy injection rates less than this value, material can build up
in the central regions of the cluster, while for larger energy
injection rates, mass is ejected from the central regions. This is
essentially the same as the explanation for mass loss from the
elliptical galaxies, except that in clusters mass is ejected from the
central regions, but still retained by the cluster as a whole. Of
course, this implicitly assumes that some of the energy injected by
the AGN is distributed over $4\pi$ steradians. This could occur
through shocks generated in the early stages of outflow and
subsequently sound waves \citep[e.g.][]{v05}, or bulk motions in the
ICM \citep[e.g.][]{heinz06a}, though these mechanisms cannot be
accounted for in the current model.

The general argument outlined above suggests that there will be a
radius, governed by the energy injection rate, within which the weight
of X-ray emitting material is governed by AGN feedback. This is
important, since by controlling the mass of gas in the central regions
of the cluster the AGN also regulates the cooling rate, and in doing
so its fuel own supply, and so on.

For clusters, the obvious region of importance is enclosed by the
cooling radius, since it contains the material that could cool within
the age of the Universe. Therefore, if AGN feedback balances the
radiative losses within the cooling radius, $\dot{E} = L_{\rm
  X}(<r_{\rm cool})$, then, according to the model, the heating should
also redistribute the material in the centre of the cluster. This
interplay should have reached a quasi-equilibrium when
\begin{eqnarray} \label{eq:int}
\dot{E} = L_{\rm X}(<r_{\rm cool}) =
-\bigg(\frac{15}{4}\bigg)\bigg(\frac{\Omega}{4\pi}\bigg)\sigma
M(r_{\rm cool})g(r_{\rm cool}) \sim \nonumber\\ -4\pi
\bigg(\frac{15}{4}\bigg)\bigg(\frac{\Omega}{4\pi}\bigg)\bigg[\frac{\nabla
    p(r_{\rm cool})}{\rho(r_{\rm cool}) }\bigg]\sigma \int_0^{r_{\rm
    cool}}r^{2}\rho(r) {\rm d}r.
\end{eqnarray}
In other words, the total mass of X-ray emitting material within
$r_{\rm cool}$, multiplied by the gravitational acceleration at
$r_{\rm cool}$ should be
\begin{equation}\label{eq:mass}
-M_{\rm T}(r_{\rm cool})g(r_{\rm cool}) =
\bigg(\frac{4}{15}\bigg)\bigg(\frac{4\pi}{\Omega}\bigg)\frac{L_{\rm
    X}(<r_{\rm cool})}{\sigma}.
\end{equation}
To be strictly correct, the AGN removes a fraction, $\Omega/4\pi$, of
this material, so the remaining weight must be $-(1-\Omega/4\pi)M_{\rm
  T}(r_{\rm cool})g(r_{\rm cool})$. 

For comparison, AGN energy injection rates can be estimated from
observations of bubble enthalpies divided by a characteristic
timescale \citep[e.g.][]{birzan,dunn05,rafferty06}. Although these
values are somewhat uncertain \citep[see][for
  example]{nusser,binney07}, it seems that more powerful AGN outbursts
do seem to be observed in more luminous clusters as well as those
systems with shorter central cooling times. Thus, AGN outbursts in
clusters are capable of injecting energy at the rate required by the
model.

Equation (\ref{eq:mass}) has been verified by examining the sample of
seven clusters discussed in \cite{pope06} and can be seen in figure 2.
The sample provides a readily available data set with all the physical
parameters required to investigate the model predictions. The main
reason for this is that the clusters are nearby so that the
temperature and the density profiles are fairly well determined to
within a few kiloparsecs of the cluster centre. Accurate temperature
and density profiles permit reasonable estimates of the total gas
mass, and gravitational acceleration, thus increasing the possibility
of detecting subtle variations due to AGN feedback. Each member of the
sample is known to host a central AGN, thus making an ideal testbed
for the model. Due to the paucity of very high X-ray luminosity
clusters, future work will focus on extending the sample predominantly
at low X-ray luminosities, to investigate and refine the model.

The best-fit in figure 2 implies a typical solid opening angle of
$\Omega/4\pi \approx 0.028$, which informed the value of the opening
angle used throughout the article. Noting that this value is twice the
effective solid opening angle of the outflow, the effective half
opening angle of a single outflow must be $\sim 14\,^{\circ}$.  This
is a factor of $\sim 3$ greater than the typical values found by radio
astronomers \citep[e.g. see][and references therein]{bintab}. Possible
explanations for this are discussed below.

In the simplest case, a buoyant bubble inflated by an AGN would expand
adiabatically to maintain pressure equilibrium with its
surroundings. According to this model, the bubble radius should behave
in the following manner \citep[e.g.][]{diehl28}
\begin{equation}
\frac{r_{\rm b}(R)}{r_{\rm b,0}} = \bigg[\frac{p_{\rm
      amb}(R)}{p_0}\bigg]^{1/3\Gamma},
\end{equation}
where $r_{\rm b}(R)$ is the bubble radius as a function of radial
displacement from its original location. $r_{\rm b,0}$ is the initial
radius of the bubble, $p_0$ is the pressure at this location, and
$p_{\rm amb}(R)$ is the ambient pressure as a function of
radius. $\Gamma$ is the polytropic index of material inside the
bubble. The half opening angle, $\theta$, can then be calculated from
${\rm tan}(\theta) = {\rm d}r_{\rm b}(R)/{\rm d}R$. Substituting
suitable values for $r_{\rm b,0}$, and $\Gamma = 4/3$ and using a
generic cluster pressure profile suggests that the adiabatic expansion
of bubbles would not result in large enough opening angles to explain
the best-fit in figure 2. In a more realistic model, \cite{pav} and
\cite{pav08} showed that the Kutta-Zhukovsky force causes bubbles to
expand further than would be expected from pressure equilibrium
arguments.

Probably the most plausible explanation is that the effective opening
angle derived from figure 2, refers to the entrainment of material
from the boundary layer between the outflow and the ambient
medium. More specifically, entraining material along the length of the
outflow, as well as through the head of the outflow. Consequently, the
outflow can collect material from within a larger solid angle than is
carved out by the main flow itself, \citep[see][for
  example]{bick94,lb02}.

Generally, $\theta$ is related to the entrainment coefficient,
$\alpha$, by $\alpha = {\rm tan}(\theta)$ \citep[e.g.][]{turner2}. In
a uniform ambient medium, and in the Boussinesq limit, the typical
entrainment coefficient for a momentum driven outflow is $\alpha \sim
0.05$, but $\alpha \sim 0.08$ for a buoyancy driven flow
\citep[see][and references therein]{turner2}. The exact values depend
slightly on whether the cross-section of the outflow is modelled as a
Gaussian or top-hat function, but the general result is that
entrainment coefficients are larger for buoyancy driven flows.

The entrainment coefficient also varies with the density contrast
between the outflow and the ambient material, and is a function of the
Mach number of the flow. \cite{deyoung94} suggests the opening angle
can be written as
\begin{equation} \label{eq:deyoung}
{\rm tan}(\theta) \sim 0.08 \bigg(\frac{\rho}{\rho_{\rm
    outflow}}\bigg)^{1/2}\bigg(\frac{1}{M}\bigg),
\end{equation}
where $\rho$ is the ambient density, $\rho_{\rm outflow}$ is the
density of the outflow, and $M$ is the Mach number of the flow
relative to the ambient material.

The best-fit from figure 2 suggests ${\rm tan}(\theta) = \alpha \sim
0.25$. Typical values for a momentum dominated outflow might be $M\sim
10$, and $\rho/\rho_{\rm outflow}\sim 10^{3}$ which would give ${\rm
  tan}(\theta) \sim 0.25$, using equation (\ref{eq:deyoung}). In the
buoyancy dominated regime, the flow speed can be described by $M \sim
1$, for which equation (\ref{eq:deyoung}) suggests $\rho/\rho_{\rm
  outflow}\sim 10$. The numbers used in this calculation are likely to
be common to all clusters, suggesting that entrainment of material
might well be what sets the effective opening angle. Furthermore, if
the outflows can be described as $\rho_{\rm outflow} w^{2} = {\rm
  constant}$, as expected for momentum conservation, then the
effective opening angle will also be constant, according to equation
(\ref{eq:deyoung}).

In figure 2, A478 appears not to be well-described by the best-fit
through the data. This may be because the best-fit assumes a uniform
effective opening angle for all the outflows in the cluster sample,
and the outflows in A478 typically have a larger effective opening
angle. There does not seem to be any particular reason to expect this
; altough A478 is the most massive cluster in this sample, it does not
exhibit a gravitational acceleration that is significantly different
in magnitude to any other clusters in the sample.

As an alternative explanation, it is interesting to note the large
temperature decrement in the cluster centre of A478 \citep[][]{sun},
which is often taken to be a signature of a strong inflow. The AGN
also appears to be injecting energy at a very low rate compared to the
radiative losses \citep[e.g. see][]{birzan}. As a result, it might be
no coincidence that A478 seems to have a larger mass within $r_{\rm
  cool}$ than would be expected from the best-fit to the data.

Table 1 lists the X-ray luminosities, weights and velocity dispersions
used to plot figure 2. For consistency with our earlier calculations
it was assumed that $\sigma^{2} = k_{\rm b}T/\mu m_{\rm p}$, where the
temperature was taken at its highest value within the cooling
radius. The values in table 1 have been rounded to two significant
figures. For this reason, and the small sample size, the uncertainties
were not calculated since they do not provide any extra information.

\begin{table*}
\centering
\begin{minipage}{140mm}
\caption{Table showing the data used to plot figure 2. Column 1 lists
  the names of the clusters, column 2 contains the observed X-ray
  luminosities within the cooling radius, $L_{\rm X}(<r_{\rm cool})$,
  column 3 gives the $M_{\rm T}(r_{\rm cool})g(r_{\rm cool})$ values,
  derived from temperature and density fits to the data. Column 4
  lists the velocity dispersions defined as $\sigma^{2} = k_{\rm
    b}T/\mu m_{\rm p}$.}
\begin{tabular}{lccc}
\hline Name & $L_{\rm X}(< r_{\rm cool})/10^{42}{\rm erg\,s^{-1}}$ & $-M_{\rm T}(r_{\rm cool})g(r_{\rm cool})/10^{37}\,{\rm
  g\,cm\,s^{-2}}$ & $\sigma/\,{\rm km\,s^{-1}}$\\ 
\hline 
Virgo & 9.8 & 0.55 & 620\\
Perseus & 670 & 20 & 1100\\
Hydra & 250 & 11 &  780\\
A2597 & 430 & 18 &  800\\
A2199 & 150 & 3.7 &  830\\
A1795 & 490 & 23 &  970\\ 
A478 & 1220 & 62 &  1100\\ 
\hline
\end{tabular}
\end{minipage}
\end{table*}\label{table1}

An alternative explanation for this correlation is not obvious. So, it
seems plausible to conclude that energy injection by AGN does reshape
the gas profiles in the central regions of clusters. As a consequence,
it is not a surprise that heating and cooling rates match relatively
well - because the systems were reconfigured into such a state by the
AGN.

\begin{figure*}
\centering
\includegraphics[width=10cm]{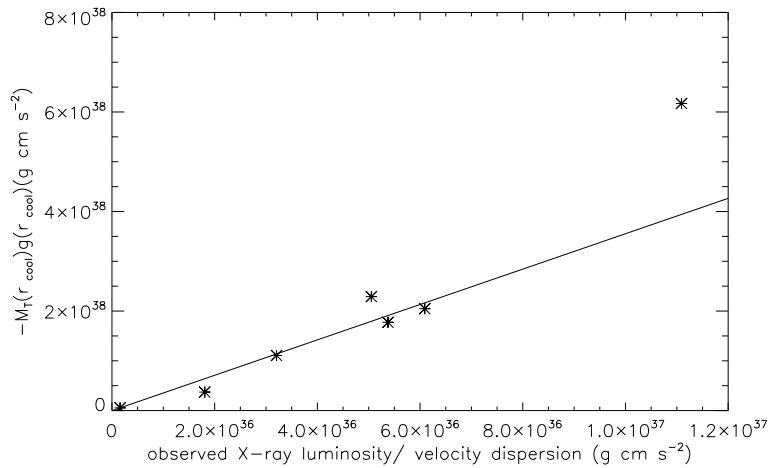}
\caption{Plot for equation (\ref{eq:mass}) using data for the Virgo,
  Perseus, Hydra, A2199, A2597, A2199, A1795, A478 taken from Pope, et
  al. (2006) and references therein. The plot confirms that $M_{\rm
    T}(r_{\rm cool})g(r_{\rm cool})$ is linearly related to the X-ray
  luminosity within the cooling, as the model predicts. The solid line
  shows the best-fit to the data, which yields an opening angle for
  the outflow of $\Omega/4\pi \sim 0.028$.}
\label{fig:flow}
\end{figure*}

Further insight into the interaction between the AGN and the ambient
material can be gained by studying equation (\ref{eq:mass}) in more
detail. If the equilibrium condition is met at all radii, then
\begin{equation} \label{eq:m2}
L_{\rm X}(<R) =
-\bigg(\frac{15}{4}\bigg)\bigg(\frac{\Omega}{4\pi}\bigg)\sigma
M(R)g(R).
\end{equation}
The X-ray luminosity is defined by $L_{\rm X} \approx \int
n^{2}\Lambda(T) {\rm d}V$, where $n$ is the electron number density of
the gas, and $\Lambda(T)$ is the temperature dependent cooling
function. For typical cluster temperatures we expect the emission to
be dominated by thermal bremsstrahlung so $\Lambda(T) \sim
T^{1/2}$. 

Differentiating both sides of equation (\ref{eq:m2}), with repsect to
volume, leaves
\begin{equation}\label{eq:tcool}
n^{2}\Lambda(T) =
-\bigg(\frac{15}{4}\bigg)\bigg(\frac{\Omega}{4\pi}\bigg)\sigma\bigg[g(R)\rho(R)+
  \frac{M_{\rm T}(R)}{4\pi R^{2}}\frac{{\rm d}g(R)}{{\rm d}R}\bigg].
\end{equation}
Factorising $g(R)\rho(R)$, we find
\begin{equation}\label{eq:tcool2}
n^{2}\Lambda(T) =
-\bigg(\frac{15}{4}\bigg)\bigg(\frac{\Omega}{4\pi}\bigg)\sigma
g(R)\rho(R)\bigg[1 + \frac{{\rm d}\ln g(R)}{{\rm d}\ln M_{\rm
      T}(R)}\bigg].
\end{equation}
The gas distribution in clusters is usually approximated by a
$\beta$-profile, which is written
\begin{equation}
\rho(R) = \frac{\rho_0}{[1 + (R/r_0)^{2}]^{\beta}},
\end{equation}
where $\rho_0$ is the central density and $r_0$ is the scale
height. Assuming the the gas temperature is independent of radius,
then $\nabla p = (k_{\rm b}T/\mu m_{\rm p}) {\rm d}\rho/{\rm d}R$, so
the gravitational accerlation is
\begin{equation}
g(R) = \frac{\nabla p (R)}{\rho(R)} = -\frac{k_{\rm b}T}{\mu m_{\rm
    p}}\frac{2\beta R}{(r_{0}^{2}+R^{2})},
\end{equation}
and equation (\ref{eq:tcool2}) can be rearranged to obtain the
critical cooling time
\begin{equation}\label{eq:tcrit}
t_{\rm cool, crit} \equiv \frac{5}{2}\frac{n k_{\rm
    b}T}{n^{2}\Lambda(T)} \sim
\bigg(\frac{4\pi}{\Omega}\bigg)\bigg(\frac{1}{3 \beta
  \sigma}\bigg)\bigg(\frac{r_{0}^{2}+R^{2}}{R}\bigg).
\end{equation}
The minimum occurs at $R = r_0$, at which point the function
simplifies to
\begin{equation}\label{eq:tcrit2}
t_{\rm cool, crit} \sim
\bigg(\frac{4\pi}{\Omega}\bigg)\bigg(\frac{1}{3
  \beta}\bigg)\bigg(\frac{r_0}{\sigma}\bigg).
\end{equation}
This relation is only approximate since it was assumed that the term
in square brackets in equation (\ref{eq:tcool2}) is $\sim 1$. However,
the approximation is reasonable because we expect ${\rm d}\ln g/{\rm
  d}\ln M_{\rm T} \sim 0$ near $r_0$.

For cooling times below this threshold, material can build up in the
core of a cluster, because the energy injection rate is insufficient
to drive it away. This build up of material is associated with the
star formation observed in clusters with short central cooling times.

As a general example, let $r_0 = 30\,{\rm kpc}$, $\sigma =
10^{3}\,{\rm km\,s^{-1}}$, $\beta=3/4$ and $\Omega/4\pi = 0.028$, so
that $t_{\rm cool, crit} \sim $ 0.5 Gyr. This agrees well with the
value found by \cite{raff08} below which star formation appears to
'switch on'. Note that this is also similar to the entropy threshold
found by \cite{cav08}.

In table 2, we list the critical cooling times calculated for the
Virgo, Perseus and Hydra clusters. These objects were chosen because
$\beta$-profiles are used to fit the gas density profiles, and the
clusters are sufficiently resolved that the fits are reliable to small
radii. $r_0$ and $\beta$ are taken from \cite{pope06}, and references
therein \citep[see][]{david01,ghizzardi04,sanders04}. The Virgo and
Perseus clusters can be described using a double $\beta$-profile
\citep[see][]{churpers,ghizzardi04} with different scale heights and
$\beta$s. For the Virgo cluster, we took the parameters for the
profile with the larger scale height, and used a single
$\beta$-profile fit for the Perseus cluster.

\begin{table*}
\centering
\begin{minipage}{140mm}
\caption{Table showing the data for the critical cooling time. Column
  1 gives the cluster name, column 2 lists the core radius of the
  density profile, column 3 gives the $\beta$ value of the density
  profile and column 4 shows the critical cooling time according to
  equation (\ref{eq:tcrit}), using the appropriate $\sigma$ values
  from table 1. The opening angle was taken to be $\Omega/4\pi =
  0.028$, from the best-fit to equation (\ref{eq:mass}). The values of
  $r_0$ and $\beta$ are taken from Pope, et al.(2006), and references
  therein.}
\begin{tabular}{lccc}
\hline 
Name & $r_{0}/\,{\rm kpc}$ & $\beta$ & $t_{\rm cool,crit}/10^{9}\,{\rm yrs}$\\ 
\hline 
Virgo & 23.3 & 0.71 & 0.5\\
Perseus & 28.5 & 0.81 & 0.3\\
Hydra & 18.6 & 0.72 &  0.4\\
\hline
\end{tabular}
\end{minipage}
\end{table*}\label{table2}

Simple analysis also suggests that the ratio $r_0/\sigma$ should be
fairly constant. According to the virial relations we expect
$\sigma^{2} \propto M/r_0 \propto r_0^{2}$, so that $r_0/\sigma$
should be approximately constant. As a result, a cooling time of $0.5$
Gyr should mark a transition in behaviour in all clusters, as appears
to be the case. Whenever, the cooling time of the gas, drops below
this critical value, we should expect to observe the consequences of
material building up in the centre of a cluster.

It has been suggested by \cite{cav08} and \cite{soker08} that the
presence of the critical cooling time is a possible indication that
central AGN are fuelled by the cold feedback mechanism
\citep[][]{pizz05}. Instead, the model presented here suggests that
the phenomenon occurs directly because of the buoyancy injected by the
AGN. This places no requirements on the AGN fuelling mechanism, except
that it must balance the radiative losses within the cooling
radius. Cold feedback still provides a possible mechanism for
achieving this.

An alternative view, presented by \cite{voit08}, is that the onset of
star formation and optical emission occur below a critical
entropy. They showed that for a central entropy greater than $K_0 \sim
30\, {\rm keV cm^{2}}$, thermal conduction transports energy at a rate
greater than can be radiated away by the hot gas, assuming that the
conduction is not heavily suppressed. In this case entropy is defined
as $K = k_{\rm b}T n_{\rm e}^{-2/3}$ \citep[e.g.][]{lloyd}, where
$k_{\rm b}$ is the Boltzmann constant, $T$ is the temperature and
$n_{\rm e}$ is the electron number density. Material below the
critical entropy would be thermally unstable, able to cool, and could
potentially fuel the central AGN. This explanation fits neatly into
the framework outlined by \cite{guo} in which non cool-core clusters
can be stabilised by thermal conduction, whereas cool core clusters
require AGN feedback and thermal conduction for stability. Their view
is not inconsistent with the model presented here. For example, the
model discussed here focusses on clusters in which AGN feedback is
important, and thermal conduction cannot balance radiative
cooling. However, complete consistency would require that AGN heating
either i) does not raise the central entropy above $\sim 30\, {\rm keV
  cm^{2}}$, otherwise the gas would never be able to cool to fuel the
AGN again, or ii) that thermal conduction is heavily suppressed. 

One difficulty with the possibility suggested by \cite{voit08} is that
the value of the thermal conduction suppression factor is highly
uncertain. In contrast, the model presented here predicts a critical
cooling time which contains no free parameters. Therefore, probably
the best way to distinguish between the models would be to determine
whether, and by how much, thermal conduction is
suppressed. Unfortunately, this is a question that cannot easily be
answered.

If the model presented here does describe some of the real mechanisms
at work in clusters, perhaps the most important aspect to highlight is
the self-tuning ability of AGN feedback. This is true in both
elliptical galaxies and clusters. The tuning is achieved by slightly
different mechanisms in each case. In elliptical galaxies, an
equilibrium is reached when the AGN has ejected so much gas that its
fuel supply becomes limited and it can no longer exceed the critical
threshold required to eject gas from the gravitational potential. In
clusters, the AGN injects energy at a particular rate given by the
details of the accretion process. This energy injection has the twin
effect of heating the gas and also moving it about within the cluster
potential. Within the radius that the AGN heating balances the
radiative losses, the heating also controls the weight of cooling
material. and therefore its own fuel supply.

\section{Summary}

We have constructed an analytical model to investigate the observable
effects of kinetic and thermal AGN feedback on elliptical galaxies and
galaxy clusters. The results show that there is a critical momentum
injection rate above which an AGN outflow will eject mass from an
elliptical galaxy. For an AGN fuelled at a rate proportional to the
classical mass cooling rate, this leads to $L_{\rm X}\propto
\sigma^{10}$ which agrees well with observations
\citep[e.g.][]{mahdavi} and reasonably with simulations
\citep[e.g.][]{bower08,dave}. If the AGN fuelling rate is only
proportional to $L_{\rm X}$ this would lead to $L_{\rm X}\propto
\sigma^{8}$. Presumeably these represent the extremes, so that a
different fuelling mechanism might lead to an intermediate scaling. 

In more massive environments, the work done by thermal energy injected
by the AGN also becomes significant. This mechanism should lead to
$L_{\rm X}\propto \sigma^{7}$, assuming the AGN fuelling rate is
proportional to the classical mass cooling rate. It is not clear
whether the observations confirm this, due to the large
scatter. Again, if the AGN fuelling rate is $ \propto L_{\rm X}$, this
would lead to $L_{\rm X}\propto \sigma^{5}$. Scalings between
$\sigma^{5}$ and $\sigma^{7}$ are also, no doubt, possible in this
regime, depending on the details of the AGN fuelling mechanism.

Above $\sigma \sim 500\,{\rm km\,s^{-1}}$ AGN heating of either sort
is unable to affect the global $L_{\rm X}-\sigma$ relation. This
corresponds to a temperature of roughly 1-2 keV and can explain the
observed break in the $L_{\rm X}-T$ relation \citep[e.g.][]{xue}. In
order for this break to be noticeable, it is required that the $L_{\rm
  X}-\sigma$ relation is steeper for ellipticals than for
clusters. Since $L_{\rm X}\propto \sigma^{4.4}$ in clusters
\citep[][]{mahdavi}, this implies that the AGN fuelling rate probably
does scales more like $L_{\rm X}/\sigma^{2}$ than $L_{\rm X}$.

In clusters of galaxies, the stronger gravity means that outflows
become buoyancy dominated above a certain length scale. Simple
estimates suggest that $\sim 20\,$kpc might be typical. It has also be
shown that buoyancy is equivalent to the work done by thermal energy
injection and since typical cooling radii in clusters are $\sim
100\,$kpc, global considerations of AGN heating in clusters can be
treated as if the energy injection is purely thermal.

It is not possible for an AGN to completely eject material from a
cluster, however, it is possible for the energy injection to
redistribute material within the central regions of the cluster. This
can lead to the self-tuning of AGN feedback so that heating roughly
matches cooling within the cooling radius. A direct consequence of
this is that the mass of hot material within the cooling radius should
be proportional to $L_{\rm X}(< r_{\rm cool})/[\sigma g(r_{\rm
    cool})]$. This appears to be true, at least for a sample of seven
nearby clusters. The only possible exception is A478, which also
displays evidence of a strong inflow and minimal AGN heating and might
therefore be expected to contain excess material in the central
regions. Furthermore, the same physical processes lead to the
derivation of a critical cooling time of $\sim 0.5$ Gyr, below which
material can pile up in the cluster centre. This is very close to the
observationally determined universal cooling time of $\sim 0.5$ Gyr,
found by \cite{raff08}, below which star formation and AGN activity
appear to be triggered.

\section{Acknowledgements}

I would like to thank the Department of Foreign Affairs and
International Trade for funding through a Government of Canada
Post-Doctoral Research Fellowship and also CITA for funding through a
National Fellowship. I also thank Arif Babul for informative
discussions and additional funding, and the anonymous referee for
positive, helpful comments that improved this work.

\bibliography{database} \bibliographystyle{mn2e}

\label{lastpage}

\end{document}